# Human Daily Activities Indexing in Videos from Wearable Cameras for Monitoring of Patients with Dementia Diseases

Svebor Karaman[1], Jenny Benois-Pineau[1], Rémi Mégret[2], Vladislavs Dovgalecs[2],
Jean-François Dartigues[3], Yann Gaëstel[3]

[1] LaBRI, Université de Bordeaux, Talence, France,
{Svebor.Karaman, Jenny.Benois}@labri.fr,
[2] IMS, Université de Bordeaux, Talence, France,
{Remi.Megret, Vladislavs.Dovgalecs}@ims-bordeaux.fr,
[3] INSERM U.897, Université Victor Segalen Bordeaux 2, Bordeaux, France,
{Jean-Francois.Dartigues, Yann.Gaestel}@isped.u-bordeaux2.fr

## Abstract

*Our research focuses on analysing human activities according to a known behaviorist scenario, in case of noisy and high dimensional collected data. The data come from the monitoring of patients with dementia diseases by wearable cameras. We define a structural model of video recordings based on a Hidden Markov Model. New spatio-temporal features, color features and localization features are proposed as observations. First results in recognition of activities are promising.*

## 1. Introduction

In the field of human behavior analysis, video brings an objective vision to the medical practitioners. Working with medical doctors, our aim is to develop a method for indexing videos via recognition of activities obtained from the monitoring of patients with wearable video cameras [12].

The daily activities of interest are specified by medical researches in the context of studies of dementia and in particular of the Alzheimer disease [5]. According to these studies the analysis of instrumental activities of daily life is one of the most important tools in the early diagnosis of Alzheimer disease. It is also mandatory in order to monitor the development of the disease.

To retrieve the structure of activities we use a Hidden Markov Model (HMM), the HMMs have been successfully applied in video analysis [3] [4]. In videos, a HMM can be used at a low level, e.g. for detecting a scene change in video [3] or at a higher level, to reveal the structure of the video according to a known grammar of events like in a tennis match [4]. Generally, HMMs are used for segmentation, classification and recognition of events in videos with a clear structure. However the wearable cameras used in patient monitoring provide a long sequence shot of daily activities. The aim of this paper is therefore to propose a scheme to divide the video into coherent segments and choose a set of descriptors efficiently describing the video content in order to identify the daily activities.

In section 2 we present the video acquisition setup and characterize the video data we use. The motion analysis and motion based temporal clustering is presented in section 3. Section 4 details the choice of descriptors and justifies the description space. In section 5 we present the scenario we have defined with medical doctors and design a HMM. Results are presented in section 6 and conclusions and perspectives of this work are drawn in section 7.

## 2. Video Acquisition Setup

### 2.1. The Device

The use of wearable cameras in the observation of patients suffering from dementia diseases can be interesting for medical doctors since it always keeps an objective vision of the patient activities contrarily to the patient's relatives who may see the situation as either worse or better than it is in reality [5]. This may lead to a late detection of the disease. Wearable cameras have been used in some previous projects like in the SenseCam project [6] where the images were recorded as a memory aid for the patient. The WearCam project [8] uses a camera strapped on the head of young children, combined with a wireless transmission from the camera to the recorder.

In our project, a mobile video recorder is used in order to obtain the best quality of image in such an embedded context. The device has to be as light as possible, since the patients are aged persons, and also to allow people to go on with their daily activities without difficulties due to the device. After studies of several positions in [12] and further in situ analysis, we have found the shoulder position to meet these objectives the best. The viewpoint obtained from this position is close to the patient's viewpoint without the problem of fixing a camera on patient's head which is uncomfortable for the patient and may lead to high non significant motion. Today our prototype uses a Fish-Eye lens with an effective diagonal angle of 150° which allows us to capture most of instrumental activities.

### 2.2. Characteristics of Recorded Videos

The video sequences shot by a wearable camera have important motion since the camera follows the movements of the person (**Fig 2**). This leads to some blur in case of strong movements (**Fig 2a**). Strong lighting changes appears when the person moves to a different room or faces a window (**Fig 2b, c**). The whole video is recorded in one shot.

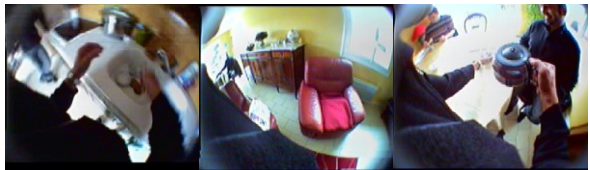

**Fig 2a.** Motion blur due to strong motion.  **Fig 2b.** High lighting while facing a window.  **Fig 2c.** High lighting while facing a window.

**Fig 2.** Example of videos acquired with a wearable camera.

## 3. Motion analysis

### 3.1. Global Motion Estimation

In wearable camera setting videos the ego-motion allows us to distinguish situations when the patient moves or remains still, e.g. sitting or standing. Therefore the motion is relevant as a descriptor of the activity of the patient. To extract this information we use the Camera Motion Detection (CMD) method previously developed in [9]. This tool estimates the motion according to a first order complete affine model. It takes as input data the motion vector $(d_{x_i}, d_{y_i})^T$ of each macro-block in the compressed video stream (**Eq. 1**). We only use the P images of MPEG videos to estimate the motion model.

$$\begin{pmatrix} dx_i \\ dy_i \end{pmatrix} = \begin{pmatrix} a_1 \\ a_4 \end{pmatrix} + \begin{pmatrix} a_2 & a_3 \\ a_5 & a_6 \end{pmatrix} \begin{pmatrix} x_i \\ y_i \end{pmatrix} \quad (1)$$

**Eq.1** Motion compensation vector, with $(x_i, y_i)$ coordinates of a block center.

### 3.2. Motion Based Temporal Segmentation

Most research in indexing video content uses videos with shot boundaries, whether cut or progressive transition [15]. In our case the whole video is a sequence shot. To define an equivalent to the traditional shot, a temporal unit, called *segment*, adapted to the specific characteristics of the video must be chosen. The typical approach nowadays consists in splitting the video into segments of a fixed duration, from half a second to one second [2], based on the latency of humans in understanding visual concepts in video. Another approach consists in more thorough use of intrinsic camera motion observed in the image plane by segmenting the sequences in so called camera view points [13]. In our case, the camera being worn by the patient, the motion is clearly related to the person's position. Therefore, defining a unique point of view as a segment is a straightforward choice.

By composing the CMD parameters (**Eq. 1**) over time, the trajectories of each corner of the image are computed. When corners trajectories reach an outbound position relatively to a predefined threshold *t*, a "cut" is detected, the current segment ends and a new one starts at the next frame. Previous experiments have shown empirically that a threshold $t = 0.2 \times$image width gives a good segmentation, meaning no viewpoint is missing and there is no over-segmentation of the video. Each segment must contain at least five frames to ensure at least one localization estimate, see **4.3**. A key frame is chosen for each segment as its temporal center.

## 4. Description Space

The video features in our case have to express the dynamics of activity, localization in the home environment and image content. Hence we propose a complex description space where low-level features such as MPEG7 Color Layout Descriptor, are merged with mid-level features in an "early fusion" way.

## 4.1. Cut histogram

A cut histogram $H_c$ is defined as a $N_c$ bins histogram, where the $i^{th}$ bin contains the number of cuts as computed by the Motion Based Temporal Segmentation in the $2^i$ previous frames. The number of bins $N_c$ has been set to 6 or 8 therefore defining a maximum temporal horizon of 256 frames i.e. 10 seconds.

A cut histogram $H_{c,seg}$ for a segment is the average of cut histograms $H_c$ of all the frames within this segment. This feature characterizes dynamics of activities related to person's displacement.

## 4.2. Translation Parameter Histogram

Translational parameters of the affine model (**Eq. 1**) are good indicators on the strength of ego-motion of a person which differs depending on the activities.

A translation parameters energy histogram $H_{tpe}$ is defined as a $N_e$ bins histogram representing the quantized probability distribution of the value of the energy of translation parameters over each video segment. Two $H_{tpe}$ histograms are built, one for each translation parameter $a_1$ and $a_4$. The ranges of the histogram bins follow a logarithmic scale defined with a step $s_h$ (**Eq. 2**) to provide a higher resolution for low-motion parameters values. This histogram is aimed at distinguishing between low and high motion activities.

$$H_{tpe}[i] += 1 \quad if \quad \log(a^2) < i*s_h \quad for \quad i=1$$
$$H_{tpe}[i] += 1 \quad if \quad (i-1)*s_h \leq \log(a^2) < i*s_h \quad for \quad i=2..N_e-1$$
$$H_{tpe}[i] += 1 \quad if \quad \log(a^2) \geq i*s_h \quad for \quad i=N_e$$

**Eq.2** Translation parameter histogram, a being either $a_1$ or $a_4$.

## 4.3. Localization

For image based localization technique we use a Bag of Features approach [11], detailed in [7] which represents each frame by a signature corresponding to a "visual word" histogram computed from SURF features [14]. The SURF descriptors are quantized using a 3 levels quantization tree [11] with a branching factor of 10, yielding 1111 dimensional signatures. The location is then estimated using 1-NN classifier.

This feature is an important discriminator with induced semantics e.g. cooking activities cannot happen in a living room. For each segment a $N_l$ bins histogram is built, $N_l$ being the number of localization classes. Each bin contains the empirical probability of being in the estimated $i^{th}$ localization class within the segment.

## 4.4. Color and Spatial Information

Colors in images are relevant to determine the environment of the patient. In our videos, the strong movements generate blur on images which does not allow us to identify details. Therefore some global information of the spatial distribution of colors is required. For this purpose we use the MPEG-7 Color Layout Descriptor (CLD) [10] which is based on the DCT transformation. Most frequently, the use of 6 coefficients for luminance and 3 coefficients for each chrominance component is depicted as a relevant choice [10]. For each segment the key frame CLD is selected as descriptor.

## 5. Hidden Markov Model

The practitioners have defined scenarios for understanding the stage of evolution of the patient's dementia. The aim is to record the patient doing daily activities where the autonomy can be evaluated. Hence, the taxonomy of states of the HMM to design is driven by these activities.

Each activity is complex so it can hardly be modeled by a single state. We consider a hierarchical HMM in which the upper level, called Activity HMM, contains states corresponding to semantic activities such as "working on a computer" or "making coffee" and where the lower level states consist of elementary states with a nested hierarchical relation: each semantic activity is modeled by an elementary HMM with $m$ states, $m$ being the global structural parameter in this two level model. The transitional matrix of the Activity HMM is fixed a priori according to the patient's home environment and all initial probabilities are set equal. The non semantic states of the Elementary HMM are modeled by Gaussian Mixture Models (GMM) in the observation space described in section 4. The Elementary HMM state transition matrix A and the observations GMM parameters are learned using the Baum-Welsh algorithm. In order to introduce the temporal regularization in the HMM, recent research focuses on segmental HMM [16]. In our work we prefer to stay in the classic HMM framework as it is lighter in terms of complexity and we obtain temporal regularization by increasing the initial looping probability of each elementary state. Previous experiments have shown an over-segmentation by the HMM when observations were used for each frame. In this work an observation corresponds to a vector obtained for a video segment by concatenating the descriptors defined in section 4. The HMMs are built using the HTK Library [1].

## 6. Results

For learning purposes, we use 10% of the total number of frames for each complex activity. These frames are used to train the localization estimator and the HMM. In this paper, we used the ground truth localization to train HMM. The other features were extracted automatically.

The tests are done over the segment observations in which we choose several subspaces of our description space. Examples of configurations are presented in the right column of table **1**. We considered the number of elementary states $m$=1,3 or 5, for each activity HMM. The initial looping probabilities $A_{ii}$ were set to 0.9. The dataset was composed of 3974 frames used for learning and 310 segments for recognition corresponding to 33 minutes of video. The 7 different activities ("moving in home office", "moving in kitchen", "going up/down the stairs", "moving outdoors", "moving in the living room", "making coffee", "working on computer") present in the video were annotated. The best recognition performances are presented in table **1**, corresponding confusion matrices are shown in **Fig 3**, lines and columns representing previously listed activities in this order.

**Table 1.** Configuration for best recognition results

| Measure | Score | Configuration |
|---|---|---|
| F-Score | 0.64 | $H_c$ + Localization<br>5 states HMMs |
| Recall | 0.7 | $H_{tpe}$ + CLD + Localization<br>3 states HMMs |
| Precision | 0.67 | $H_c$ + Localization<br>5 states HMMs |

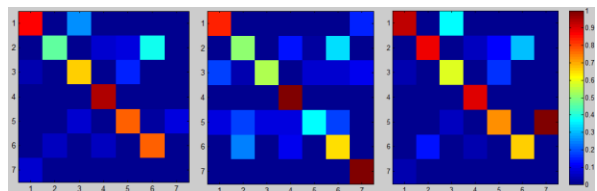

**Fig 3a.** F-Score    **Fig 3b.** Recall    **Fig 3c.** Precision

**Fig 3**. Confusion matrices for best recognition results

The results are very good for some activities such as "moving in home office" with precision of 0.94, recall of 0.81 and F-Score of 0.87 in the configuration ($H_c$ + Localization, 5 states HMMs). However, other activities are much more difficult to detect, such as "moving in the kitchen" with a F-Score of 0.47. This reveals that the visual description space is still limited to distinguish activities which take place in the same environment and involve similar motion activity. The activity "working on computer" is also specific because the temporal segmentation gives only 2 segments in the video even if this activity represents thousands of frames. The temporal segmentation has to be refined.

## 7. Conclusions and Perspectives

This article has presented a human activity indexing method based on HMM with a mixed description space. Results show that the activities which have a strong correlation with localization are well identified. Nevertheless, it is necessary to enrich the description space by features describing objects, audio content and patient's behavior (e.g. manual activity). Due to the diversity of home environments, it would be hardly possible to train generic models. Hence, we will have to define efficient protocols for large-scale training.